\documentclass[reprint,amsmath,amssymb,aps,pre,showpacs,eqsecnum]{revtex4-1}

\usepackage{graphicx}      % Include figure files
\usepackage{dcolumn}       % Align table columns on decimal point
\newcolumntype{d}[1]{D{.}{.}{#1}}

    \usepackage[usenames]{color}

\def\bal#1\eal{\begin{align}#1\end{align}}

\newcommand{\zz}{\hat{\zeta}}

\newcommand{\sm}{\mathsf}

\newcommand{\beq}{\begin{equation}}
\newcommand{\eeq}{\end{equation}}
\newcommand{\beqn}{\begin{eqnarray}}
\newcommand{\eeqn}{\end{eqnarray}}

\newcommand\beqa{\begin{eqnarray}}
\newcommand\eeqa{\end{eqnarray}}
\newcommand{\nn}{\nonumber\\}
\newcommand{\dd}{\text{d}}

\newcommand{\ii}{\text{i}}

\newcommand{\ed}{\end{document}}

\begin{document}
\title{Equation of state for five-dimensional hyperspheres from the chemical-potential route}

\author{Ren\'e D. Rohrmann}
\email{rohr@icate-conicet.gob.ar}
\homepage{http://icate-conicet.gob.ar/rohrmann}
\affiliation{Instituto de Ciencias Astron\'omicas, de la Tierra y del
Espacio (ICATE-CONICET), Avenida Espa\~na 1512 Sur, 5400 San Juan, Argentina}

\author{Andr\'es Santos}
\email{andres@unex.es}
\homepage{http://www.unex.es/eweb/fisteor/andres/}

\affiliation{Departamento de F\'{\i}sica and Instituto de Computaci\'on Cient\'ifica Avanzada (ICCAEx), Universidad de
Extremadura, Badajoz, E-06071, Spain}

\date{\today}
\begin{abstract}
We use the Percus--Yevick approach in the chemical-potential route to evaluate the equation of state of hard hyperspheres in five dimensions.
The evaluation requires the derivation of an analytical expression for the contact value of the pair distribution function between particles of the bulk fluid and a solute particle with arbitrary size.
The equation of state is compared with those obtained from the conventional virial and compressibility thermodynamic routes and the associated virial coefficients  are computed. The pressure calculated from all routes is exact up to third density order, but it deviates with respect to simulation data as density increases, the compressibility and the chemical-potential routes exhibiting smaller deviations than the virial route. Accurate linear interpolations between the compressibility route and either the chemical-potential route or the virial one are constructed.
\end{abstract}

%\pacs{02.50.-r, 05.20.Jj, 51.30.+i} % PACS, the Physics and Astronomy
                                    % Classification Scheme.
      %02.50.-r Probability theory, stochastic processes and statistics
      %05.20.Jj Statistical mechanics of classical fluids
      %34.20.Cf Interatomic potentials and forces
      %51.30.+i Thermodynamic properties, equations of state
      %52.25.Jm Ionization of plasmas

%Use showkeys class option if keyword display desired

\pacs{
61.20.Gy,       %Theory and models of liquid structure
05.70.Ce, 	%Thermodynamic functions and equations of state
61.20.Ne, 	%Structure of simple liquids
65.20.Jk 	%Studies of thermodynamic properties of specific liquids
}

\maketitle
%%%%%%%%%%%%%%%%%%%%%%%%%%%%%%%%%%%%%%%%%%%%%%%%%%%%%
\section{ Introduction} \label{s.intr}

The thermodynamic and equilibrium statistical-mechanical behavior of liquids and dense gases can be conveniently described by means of distribution functions \cite{H56,BH76,R09,HM06}. The most widely used one is the pair correlation function or radial distribution function (RDF) $g(r)$, which is particularly appropriate for the study of systems of particles interacting with a pairwise additive potential. For such fluids, there exists a set of well-established, rigorous relationships connecting thermodynamic quantities with configuration integrals over RDF. In particular, the compressibility, energy, and pressure (or virial) equations provide well-known routes to the thermodynamic properties of the fluid \cite{H56,BH76, R09, HM06,S14}.

Thermodynamic properties can also  be obtained from another route that
connects $g(r)$ with the chemical potential $\mu$ through the charging process of a test particle in the fluid \cite{O33, K35, H56}. This represents the chemical-potential route ($\mu$ route), which has remained almost unexplored until recent years, when it was used to obtain a new Percus--Yevick (PY) equation of state (EOS) for the hard-sphere system \cite{S12b}. Subsequently, this has been formally generalized to arbitrary multicomponent systems and applied to additive hard-sphere (AHS) mixtures \cite{SR13}.
In addition, the EOS of sticky hard spheres (or Baxter model \cite{B68}) was derived and the critical point associated with a liquid-gas transition was captured by PY results in this way \cite{RS14}.

All these routes (virial, energy, compressibility, and $\mu$) are formally exact and thermodynamically equivalent, provided that the exact RDF is employed \cite{S14}. Actually, however,  only approximate evaluations of $g(r)$ are available for systems in dimensions $d>1$ with nontrivial interactions. In general, the RDF is obtained with some accuracy from numerical simulations \cite{HM06}, by solving integral equation approximations (e.g., the PY \cite{PY58} and hypernetted chain approaches \cite{vLGB59,M60}) or from density functional theories \cite{E79,R89}. In this context, the $\mu$ route has demonstrated to provide an alternative and useful path for the study of the thermodynamic properties of fluids.

With the aim of  extending the use of the $\mu$ route and to gain insight into its properties, in this paper we apply the $\mu$ method to find the EOS of the hard hypersphere fluid at spatial dimension $d=5$. Besides the intrinsically interesting properties of hard particle systems, they are an important basis for constructing more complicate models, so that there are active theoretical efforts to study them in dimensions $d>3$ (see, for instance, Refs.\ \cite{CB86,LM90,SYH01,FSL02,GAH01,CM04a,L05,CM06,TS06a,TS06,RS07,AKV08,vMCFC09,LBW10,LS11,PSK11,ER11,ZP14} and references therein).

The $\mu$ route is based on a charging process where a test particle (solute) with tunable interaction is inserted into the bulk fluid (solvent particles). As a consequence, the application of the method requires the solute-solvent RDF of the corresponding binary system in the infinite dilution limit.
To our knowledge, the only systematic theoretical method for the evaluation of the RDF of AHS fluid mixtures at dimensions higher than $d=3$ is the so-called rational-function approximation (RFA) \cite{RS11}. Its simplest implementation provides the solution of the Ornstein--Zernike relation coupled with the PY closure for AHS mixtures in odd dimensionalities \cite{RS11b}.
Therefore, we adopt the RFA technique at the PY level in the present study.

This work is organized as follows.
In Sec.\ \ref{s.frame} we present the basic formulation of thermodynamics routes for hypersphere systems.  Section \ref{s.RFA} gives, within the PY theory,
the five-dimensional RDF for a coupled particle with arbitrary size, a key quantity for evaluating the EOS in the $\mu$ route. The technical details are elaborated in the Appendixes.
The results are presented in Sec.\ \ref{s.res}.
Finally, in Sec.\ \ref{s.fr} we offer some conclusions.

%%%%%%%%%%%%%%%%%%%%%%%%%%%%%%%%%%%%%%%%%%%%%%%%%%%%%
\section{Framework} \label{s.frame}

\begin{figure}
\includegraphics[width=8cm]{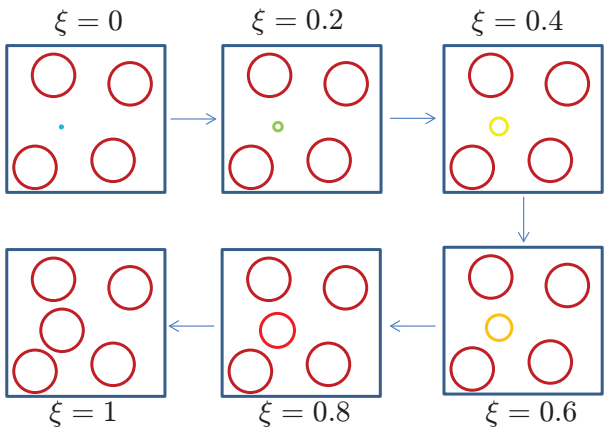}
\caption{(Color online) Cartoon of the charging process of the solute from a point particle ($\xi=0$) to a particle equivalent to any solvent particle ($\xi=1$).}
\label{sketch}
\end{figure}

The EOS for a single component system of hard $d$-sphere particles can be expressed as a relationship between the compressibility factor $Z=P/\rho k_{\rm B}T$ (where $P$ is the pressure, $\rho$ is the average particle density, $k_{\rm B}$ is the Boltzmann constant, and $T$ is the temperature) and the packing fraction $\eta= v_d \rho \sigma^d$, with  $\sigma$ the diameter of the particles and $v_{d}=(\pi/4)^{d/2}/\Gamma(1+d/2)$ the volume of a $d$-dimensional sphere of unit diameter.
In the $\mu$ route, the EOS is given by \cite{S12b,SR13,S14,BS14}
\beq  \label{zmu}
Z_\mu(\eta)=-\frac{\ln(1-\eta)}{\eta} +{(2^{d}-1)\eta
\left[{\overline{g}}(\eta) -\int_0^1 \dd t \,t {\overline{g}}(\eta t) \right]},
\eeq
with
\beq  \label{zmuf}
{\overline{g}(\eta)\equiv\frac{d}{2^d-1}\int_0^1\dd\xi \,\left(1+\xi\right)^{d-1} g(\eta;\xi)},
\eeq
where $g(\eta;\xi) \equiv g_{12}(\sigma_{12}^+)$ is the  contact value of the solute-solvent RDF, which depends on both $\eta$ and the coupling parameter $\xi$. The latter is defined as the ratio between the solute and the solvent diameters, so that  the minimum possible distance between solute and solvent particles is
\beq
\sigma_{12}=\frac{1+\xi}{2} \sigma.
\eeq
Thus, $\xi$ regulates the strength of the interaction between the test particle and the rest of the fluid. When $\xi=0$ the test particle is a point that cannot penetrate the solvent particles, while
when $\xi=1$ the test particle is  indistinguishable from any particle of the bulk fluid. This charging process is schematically illustrated by Fig.\ \ref{sketch}.

For comparison, the compressibility factor in the virial and compressibility routes are expressed by \cite{RS07}
\beq \label{Zv}
Z_{\rm v}(\eta)=1+2^{d-1}\eta g(\eta;1),
\eeq
and
\beq \label{Zc}
Z_{\rm c}(\eta)=\int_0^1 \dd t\, \chi^{-1}(\eta t),
\eeq
where $\chi(\eta)$ is the isothermal susceptibility. Since hard spheres are athermal, the energy route to the EOS becomes useless.

In order to evaluate $g_{12}(\sigma_{12}^+)$ within the RFA approach \cite{RS11}, it is convenient to introduce the Laplace functional defined by
\beq \label{Gdef}
G_{ij}(s)= \int_0^\infty \dd r\, r g_{ij}(r) \theta_n(sr) e^{-sr},
\eeq
where $g_{ij}(r)$ is the RDF of the pair ($i,j$) and
$\theta_n(sr)$ is the reverse Bessel polynomial of order $n=(d-3)/2$ \cite{RS07}. This functional is directly related to the static structure factors
$S_{ij}(k)$ of a multicomponent fluid,
\beq \label{Sijh}
S_{ij}(k)=x_i \delta_{ij}+  (2\pi)^{(d-1)/2} \rho x_i x_j \ii
 \frac{G_{ij}(\ii k)-G_{ij}(-\ii k)}{k^{d-2}},
\eeq
where $k$ is the wavenumber, $x_i$ is the mole fraction of species $i$, and $\ii$ is the imaginary unit.
The functional $G_{ij}(s)$ provides us with all the necessary information about the structure and thermodynamics  of the fluid state. In particular,  the contact values are \cite{RS11}
\beq \label{gG}
\sigma_{ij}^{(d-1)/2} g_{ij}(\sigma_{ij}^+) = \lim_{s\rightarrow\infty}{s^{(5-d)/2}}
e^{\sigma_{ij}s}G_{ij}(s),
\eeq
$\sigma_{ij}$ being the contact distance of the pair ($i,j$).

%%%%%%%%%%%%%%%%%%%%%%%%%%%%%%%%%%%%%%%%%%%%%%%%%%%%%
\section{PY approach} \label{s.RFA}

\begin{figure}
\includegraphics[width=8cm]{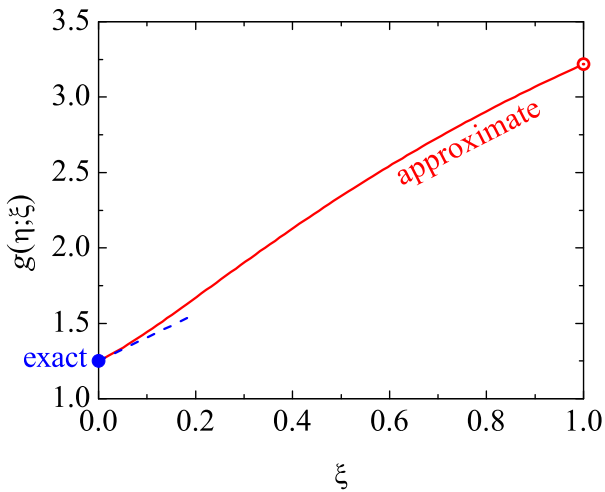}
\caption{(Color online) Plot of the solute-solvent contact value $g(\eta;\xi)$ versus the solute-to-solvent size ratio $\xi$ at a packing fraction $\eta=0.2$. The values of $g(\eta;\xi)$ and its slope at $\xi=0$ are exact, but the full curve is the approximate prediction of the PY theory [see Eq.\ \protect\eqref{g12c}].}
\label{fig1}
\end{figure}

The exact solution of the PY approximation for $d$-odd AHS mixtures with any number of components has been described in detail in Ref.\ \cite{RS11} within the framework of the RFA method. For the sake of consistency, the main expressions particularized to Five-dimensional ($5{\rm D}$)  binary mixtures are presented in Appendix \ref{appA}.
{}From them one can take the limit of solute infinite dilution to obtain the solute-solvent contact value $g(\eta;\xi)$. The details are worked out in Appendix \ref{appB} and here we only quote the final result:
\beq
{g(\eta;\xi)=\frac{2+3\eta-\zeta+3(1-\eta+\zeta \xi)\xi+\zeta^2\xi^3/(1-\eta)}{(1+\xi)^2(2+3\eta-\zeta)(1-\eta+\zeta\xi)}},
\label{g12c}
\eeq
where
\beq
\label{a.xi}
\zeta=\sqrt{1+18\eta+6\eta^2}.
\eeq
In the special case $\xi=1$ one recovers the solvent-solvent value, namely.
\beq
{g(\eta;1)=\frac{\zeta^3-1+33\eta+87\eta^2+6\eta^3}{60\eta(1-\eta)^3}}.
\label{g11c}
\eeq
In the opposite limit $\xi=0$ one obtains the \emph{exact} results \cite{S12b} $g(\eta;0)=1/(1-\eta)$ and {$\left.\partial g(\eta;\xi)/\partial \xi\right|_{\xi=0}=5\eta/(1-\eta)^2$}.
Figure \ref{fig1} shows the $\xi$-dependence of $g(\eta;\xi)$ for the representative packing fraction $\eta=0.2$.

Once $g(\eta;\xi)$ has been analytically determined, the integral in Eq.\ \eqref{zmuf} can be evaluated with the explicit result
\bal
{\overline{g}}(\eta)=&\frac{{5}}{{62}\zz^3(2+3\eta-\zeta)}\Big[\frac{2+3\eta}{1-\eta}\zz(5\zz-2)+2
\nn
&-5\zz+\frac{20}{3}\zz^2+\frac{2}{3}\zz^3+\frac{31}{15}\zz^4+2(1-\zz)^2\nn
&\times\left(\frac{2+3\eta}{1-\eta}-\zz-\zz^{-1}\right)\ln(1+\zz)\Big],
\label{f(eta)}
\eal
where $\zz\equiv \zeta/(1-\eta)$.
The compressibility factor in the $\mu$ route can be easily evaluated from Eq.\ \eqref{zmu} by  numerical integration.

Equation \eqref{g11c} can be used to evaluate the virial route to the EOS from Eq.\ (\ref{Zv}).
Finally, the isothermal susceptibility of $5{\rm D}$ hyperspheres in the PY approach is \cite{FI81,RS07}
\beq
\chi(\eta)=\frac{{(1-\eta)^2}}{\zeta^2}[5(1+6\eta+3\eta^2)-2(2+3\eta){\zeta}].
\eeq
{}From here one can obtain the compressibility route to the EOS via Eq.\ (\ref{Zc}) as
\beq
{Z_{\text{c}}=\frac{2 \zeta^5 - 2 + 135 \eta + 1230 \eta^2 + 3645 \eta^3 + 990 \eta^4 +
 252 \eta^5}{225 \eta (1 - \eta)^5}}.
 \label{ZZc}
\eeq

%%%%%%%%%%%%%%%%%%%%%%%%%%%%%%%%%%%%%%%%
\section{Results} \label{s.res}

\subsection{Virial coefficients}
We start this section by considering the virial expansion
\beq
Z(\eta)= 1+\sum_{n=2}^\infty b_n \eta^{n-1}
\eeq
for the $5\text{D}$ fluid as predicted by the three routes in the PY approximation. The virial coefficients $b_n^{(\text{v})}$ and $b_n^{(\text{c})}$ corresponding to the virial and compressibility routes are obtained from Eqs.\ \eqref{g11c} [combined with Eq.\ \eqref{Zv}] and \eqref{ZZc}, respectively.
As for the $\mu$ route, Eq.\ \eqref{zmu} implies that{
\beq
{b_n^{(\mu)}=\frac{1+31(n-1)\overline{g}_{n-2}}{n}},
\eeq
where the coefficients ${\overline{g}_n}$, which are defined by the expansion
\beq
{\overline{g}(\eta)=\sum_{n=0}^\infty \overline{g}_n \eta^n},
\eeq
are easily obtained from Eq.\ \eqref{f(eta)}.}

{\renewcommand{\arraystretch}{2}
\begin{table*} [t]%[htb!]
\caption{First $12$ virial coefficients obtained from the PY theory according to the virial ($b_n^{(\text{v})}$), compressibility ($b_n^{(\text{c})}$), and $\mu$  ($b_n^{(\mu)}$) routes.} \label{t.b}
\begin{ruledtabular}
\begin{tabular}{cccc}
$n$     &  $b_n^{(\text{v})}$ &  $b_n^{(\text{c})}$ & $b_n^{(\mu)}$ \\
\hline
$2$  & $16$ &  $16$ & $16$ \\
$3$  & $106$ &  $106$ & $106$ \\
$4$  & $196$ &  $\displaystyle{\frac{1459}{4}}$ & $\displaystyle{\frac{3561}{16}}$ \\
$5$  & $\displaystyle{\frac{1697}{2}}$ &  $\displaystyle{\frac{2147}{2}}$ & $\displaystyle{2000\ln 2-\frac{1607}{3}}$ \\
$6$  & $-2999$  & $\displaystyle{\frac{12\,233}{8}}$   & $\displaystyle{-\frac{303\,125}{3}\ln 2+\frac{19\,465\,513}{288}}$ \\
$7$  & $\displaystyle{\frac{164\,989}{4}}$ &  $10\,591$  &  $\displaystyle{\frac{47\;221\;875}{14}\ln 2-\frac{64\,427\,077}{28}}$ \\
$8$  & $-466\,319$ & $\displaystyle{-\frac{3\,790\,541}{64}}$ &   $\displaystyle{-\frac{2\,998\,209\,375}{32}\ln 2+\frac{49\,556\,094\,191}{768}}$ \\
$9$  & $\displaystyle{\frac{184\,953\,797}{32}}$ &  $\displaystyle{\frac{24\,293\,447}{32}}$ &  $\displaystyle{2\,345\,096\,875\ln 2-\frac{87\,495\,533\,291}{54}}$ \\
$10$ & $\displaystyle{-\frac{1\,193\,474\,849}{16}}$  & $\displaystyle{-\frac{1\,124\,833\,117}{128}}$  &
$\displaystyle{-\frac{1\,756\,815\,571\,125}{32}\ln 2+\frac{19\,449\,143\,196\,527}{512}}$ \\
$11$ & $\displaystyle{\frac{63\,809\,313\,739}{64}}$  & $\displaystyle{\frac{3\,463\,610\,957}{32}}$  &
$\displaystyle{\frac{431\,924\,172\,778\,125}{352}\ln 2-\frac{224\,301\,227\,394\,011}{264}}$ \\
$12$ & $\displaystyle{-\frac{109\,613\,124\,547}{8}}$  & $\displaystyle{-\frac{704\,051\,293\,633}{512}}$  &
$\displaystyle{-\frac{6\,786\,922\,855\,865\,625}{256}\ln 2+\frac{2\,369\,375\,540\,853\,543\,359}{129\,024}}$ \\
 \end{tabular}
 \end{ruledtabular}
 \end{table*}
 }
%%%%%%%%%%%%%%%%%%%%%%%%%%%%%%%%%%%%%%%%%%%%%%%%%%%%%%%%%%%%%%%%%%%%%%%

%%%%%%%%%%%%%%%%%%%%%%%%%%%%%%%%%%%%%%%%%%%%%%%%%%%%%%%%%%%%%%%%%%%%%%%
\begin{table} [t]%[htb!]
\caption{Comparison between the numerical values of the virial coefficients for $4\le n\le 12$ as obtained from  several routes in the PY approximation and the exact values ($b_n^{(\text{ex})}$) for $n=4$ \protect\cite{L05} and $5\leq n\leq 12$ \protect\cite{ZP14}.}
\label{t.bnum}
\begin{ruledtabular}
\begin{tabular}{cllll}
  $n$ & $10^4 b_n^{(\text{ex})}/b_2^{n-1}$&  $10^4 b_n^{(\text{v})}/b_2^{n-1}$ &  $10^4b_n^{(\text{c})}/b_2^{n-1}$ & $10^4b_n^{(\mu)}/b_2^{n-1}$\\
\hline
   $4$  &      $759.7248$ &       $478.5156$ &       $890.5029$ &       $543.3655$ \\
   $5$  &      $129.5219(16)$ &      $129.4708$ &      $163.8031$ &      $ 129.7955$ \\
   $6$  &     $9.8184(19)$ &     $-28.6007$ &      $14.5829$ &     $-23.3475$ \\
   $7$  &    $4.165(2)$ &    $ 24.585$ &     $6.313$ &    $ 22.061$ \\
   $8$  &   $-1.127(4)$ &   $-17.372$ &    $-2.206$ &   $-15.557$ \\
   $9$  &  $0.789(5)$ &  $13.457$ &   $1.768$ &  $ 12.130$ \\
  $10$  & $-0.468(10)$ & $-10.855$ &  $-1.279$ & $-9.824$ \\
  $11$  & $0.309(11)$& $9.068$ &  $0.984$ & $8.235$ \\
  $12$  & $-0.23(2)$& $-7.79$ &  $-0.78$ & $-7.09$ \\
\end{tabular}
 \end{ruledtabular}
\end{table}
%%%%%%%%%%%%%%%%%%%%%%%%%%%%%%%%%%%%%%%%%%%%%%%%%%%%%%%%%%%%%%%%%%%%%%%

% =========================== figure
\begin{figure}
{\includegraphics[width=8cm]{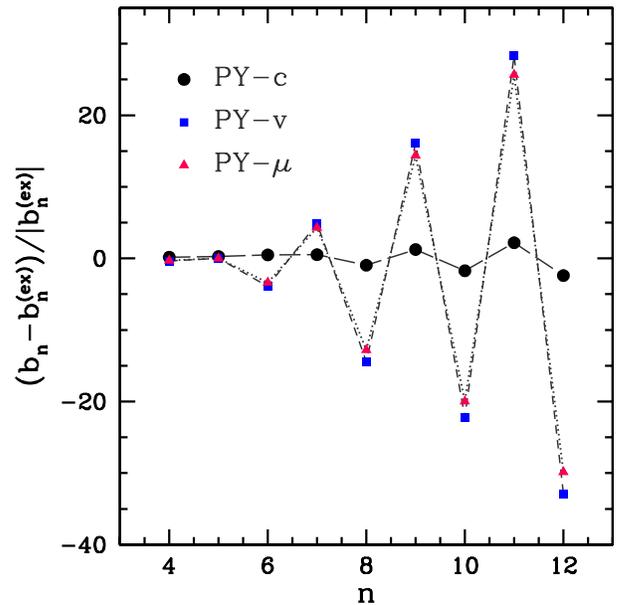}}
  \caption{(Color online) Normalized differences between the virial coefficients in the virial (squares), compressibility (circles) and $\mu$ (triangles) routes, with respect to the exact values \protect\cite{L05,ZP14}.}
\label{f.vi}
\end{figure}
% =========================================

Table  \ref{t.b} contains the first $12$ virial coefficients obtained from the three routes in the PY theory.
As is well known, the PY theory yields the exact virial coefficients up to third order, regardless of the thermodynamical route.
However, discrepancies among results from different routes appear for upper order coefficients ($n\ge4$).
As a peculiar feature, it may be observed that, in contrast to the three-dimensional (3D) case \cite{S12b}, the $\mu$ route yields irrational virial coefficients (due to the logarithmic term) for $n\ge 5$. On the other hand, the virial and compressibility routes yield rational numbers for the virial coefficients of hypersphere systems in all odd dimensions \cite{RRHS08}.

Table \ref{t.bnum} shows the three sets of PY virial coefficients in numerical format and compares them with {the exact result for $n=4$ \cite{L05} and with  recent accurate values  for $5\leq n\leq 12$ \cite{ZP14}}.
The relative deviations are displayed in Fig.\ \ref{f.vi}.
It is quite clear that the compressibility route gives the closest agreement with the exact values, while the
 values calculated from the $\mu$ and virial routes are rather similar, with a slight improvement of  $b_n^{(\mu)}$ over $b_n^{(\text{v})}$.
It is interesting to remark that the three PY routes capture the alternating sign change between $n=7$ and $n=12$, while a  negative sign of $b_6$ is wrongly anticipated by the virial and $\mu$ routes. In the PY case, the alternating character is related to the existence of a branch point singularity on the negative real axis (at $\eta=-3/2+5\sqrt{3}/6\simeq -0.0566$), which determines the radius of convergence of the virial series \cite{RRHS08}.

Regarding the closeness between $b_n^{(\mu)}$ and  $b_n^{(\text{v})}$, it turns out to be  higher in five dimensions than in three dimensions. While in the 3D case the ratio $b_n^{(\text{v})}/b_n^{(\mu)}$ monotonically decreases from $0.955$ ($n=4$) to $\frac{4}{5}$ ($n\to\infty$) \cite{S12b}, in the 5D case the ratio first increases from $0.881$ ($n=4$) to a maximum value $1.225$ ($n=6$) and then (from $n=8$)  monotonically decreases towards  an asymptotic value $1.0646$. It can then be speculated that the general similarity between  $b_n^{(\mu)}$ and  $b_n^{(\text{v})}$ tends to increase with increasing dimensionality.
{We will return to this point at the end of Sec.\ \ref{s.resB}.}

\subsection{Equation of state at finite density}
\label{s.resB}

As said before, the compressibility factor of $5{\rm D}$ hyperspheres in the PY $\mu$ route can be obtained through numerical  integration from Eq.\ (\ref{zmu}) by making use of the explicit expression \eqref{f(eta)}. The virial and compressibility routes yield Eqs.\ \eqref{Zv}, combined with Eq.\ \eqref{g11c}, and \eqref{ZZc}, respectively.

% =========================== figure
\begin{figure}
\includegraphics[width=8cm]{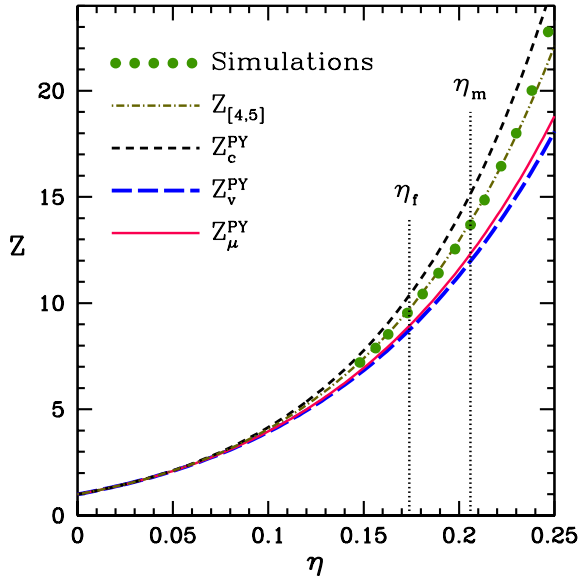}
  \caption{(Color online) Compressibility factors evaluated from computer simulations \protect\cite{LBW10} (symbols) and from an accurate Pad\'e approximant $Z_{[4,5]}$ (dash-dotted line) \protect\cite{BW05,LBW10} are compared with PY results in the virial (long dashed line), compressibility (short dashed line), and $\mu$ (solid line) routes. The dotted vertical lines mark the freezing and melting packing fractions \protect\cite{vMCFC09}.
\label{f.Z}}
\end{figure}
%==============================================

Figure \ref{f.Z} shows the density dependence of $Z(\eta)$ according to the PY theory in the virial ($Z_\text{v}^{\rm PY}$), compressibility ($Z_\text{c}^{\rm PY}$),  and $\mu$ ($Z_\mu^{\rm PY}$) routes. Predictions of computer simulations  \cite{LBW10} are denoted by symbols, while the dash-dotted line represents a $Z_{[4,5]}$ Pad\'e approximant \cite{BW05,LBW10} based on high precision calculations of the first ten virial coefficients \cite{CM06}.
The freezing ($\rho\sigma^5\approx 1.06$, $\eta\approx0.174$) and melting ($\rho\sigma^5\approx 1.25$, $\eta\approx0.206$) densities calculated for the $D_5$ lattice \cite{vMCFC09} are also indicated. As  expected,
all three PY routes  converge to the exact results at very low density. On the other hand, as the packing is increased, the compressibility route overestimates $Z$, whereas the virial and $\mu$ routes underestimate it, the values from the $\mu$ route being slightly better than those from the virial route.

 % =========================== figure
\begin{figure}
\includegraphics[width=8cm]{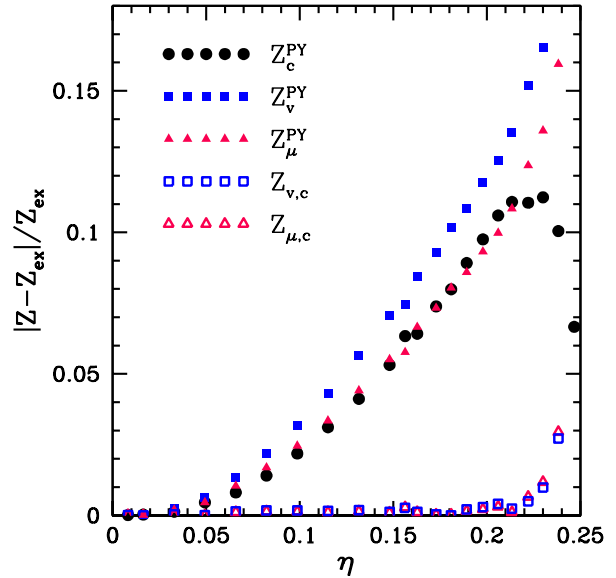}
  \caption{(Color online) Relative differences between the compressibility factor $Z_{\text{ex}}$ evaluated from computer simulations ($\eta>0.14$) \cite{LBW10} or from the Pad\'e approximant $Z_{[4,5]}$ ($\eta<0.14$) \cite{BW05,LBW10} and the PY results in several routes (filled symbols). Open symbols correspond to the interpolations $Z_{\rm v,c}$ (squares) and $Z_{\rm \mu,c}$ (triangles) from Eqs.\ (\ref{Zvc}) and (\ref{Zmc}), with $\alpha=0.44$ and $\alpha=\frac{1}{2}$, respectively.}
\label{f.Zs}
\end{figure}
%==============================================

The relative deviations of $Z_\text{v}^{\rm PY}$, $Z_\text{c}^{\rm PY}$,  and $Z_\mu^{\rm PY}$ from the exact compressibility factor $Z_{\text{ex}}$ are plotted in Fig.\ \ref{f.Zs}.
It can be observed  that the deviations of the compressibility and $\mu$ routes from computer simulations  are very similar (but with different sign) within the fluid phase ($\eta<0.2$).
For instance, at $\eta=0.1$, the deviations between simulation or Pad\'e-approximant results and the PY theory become $3.2$\%, $2.2$\%, and $2.5$\% for the virial, compressibility, and $\mu$ routes, respectively, while they increase to $11.7$\%, $9.8$\%, and $9.3$\%, respectively, when the packing fraction $\eta=0.2$ is reached.

%
% =========================== figure
\begin{figure}
\includegraphics[width=8cm]{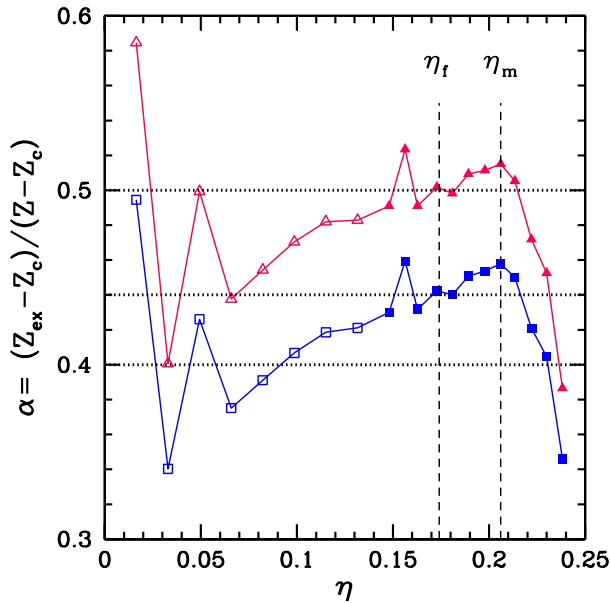}
  \caption{(Color online) Ratio $\alpha=(Z_{\text{c}}^{\text{PY}}-Z_{\text{ex}})/(Z_{\text{c}}^{\text{PY}}-Z)$  with $Z=Z_{\text{v}}^{\text{PY}}$  (squares) and $Z=Z_{\mu}^{\text{PY}}$ (triangles), using the compressibility factor $Z_{\text{ex}}$ obtained either by simulations \cite{LBW10} (filled symbols) or from the Pad\'e approximant  $Z_{[4,5]}$ \cite{BW05,LBW10}. As reference, horizontal lines denote the values $\alpha=\frac{1}{2}$, {$\alpha=0.44$}, and $\alpha=\frac{2}{5}$, and vertical lines refer to the freezing and melting packing fractions \cite{vMCFC09}.}
\label{f.alfa}
\end{figure}
% =========================================

Looking for a better agreement with numerical experiments from PY solutions, one can construct new EOSs through Carnahan--Starling-like interpolations of the form
\beq \label{Zvc}
Z_{\rm v,c}(\eta)=\alpha Z_{\rm v}^{\rm PY}(\eta)
                 +(1-\alpha)Z_{\rm c}^{\rm PY}(\eta),
\eeq
\beq \label{Zmc}
Z_{\rm \mu,c}(\eta)=\alpha Z_\mu^{\rm PY}(\eta)
                   +(1-\alpha)Z_{\rm c}^{\rm PY}(\eta).
\eeq
To test this possibility, one can define density-dependent weight functions $\alpha_{\text{v,c}}(\eta)=[Z_{\text{c}}^{\text{PY}}(\eta)-Z_{\text{ex}}(\eta)]/[Z_{\text{c}}^{\text{PY}}(\eta)-Z_{\text{v}}^{\text{PY}}(\eta)]$
and $\alpha_{\mu,\text{c}}(\eta)=[Z_{\text{c}}^{\text{PY}}(\eta)-Z_{\text{ex}}(\eta)]/[Z_{\text{c}}^{\text{PY}}(\eta)-Z_{\mu}^\text{{PY}}(\eta)]$ to check to which extent they are weakly dependent on $\eta$. This is done in
Fig.\ \ref{f.alfa}, which shows evaluations of the parameter $\alpha(\eta)$ from available data. Variations observed  at the lowest densities ($\eta<0.07$) are due to the ratio of very small values; actually, all three routes yield accurate results for very dilute gas conditions. On the other hand, it is clear from Fig.\ \ref{f.alfa} than there is not a unique value of $\alpha_{\text{v,c}}(\eta)$ or $\alpha_{\mu,\text{c}}(\eta)$ in the fluid phase region ($\eta<0.2$). However, it is to be noted that $\alpha=\frac{1}{2}$ and $\alpha=0.44$ may be  reasonable choices for $Z_{\rm \mu,c}$ and  $Z_{\rm v,c}$, respectively. It can be observed from Fig.\ \ref{f.alfa} that the value $\alpha=0.44$ in Eq.\ \eqref{Zvc} is better in the region $\eta\approx 0.2$ than the simpler value $\alpha=\frac{2}{5}$ proposed in Ref.\ \cite{S00}.
Data obtained from the hybrid compressibility factors $Z_{\rm v,c}$ (with $\alpha=0.44$) and $Z_{\rm \mu,c}$ (with $\alpha=\frac{1}{2}$) are also included in Fig.\ \ref{f.Zs}. Both EOSs have discrepancies  with respect to  simulation results lower than $0.3$\%  for $\eta<0.2$.

The improvement of the $\mu$ route over  the virial route observed in the virial coefficients (see Table \ref{t.bnum} and Fig.\ \ref{f.vi}) as well as in the EOS at finite densities (see Figs.\ \ref{f.Z} and \ref{f.Zs}) can be rationalized by the following heuristic argument \cite{S12b}.
Comparison between the exact statistical-mechanical formulas \eqref{zmu} and \eqref{Zv} shows that, while the $\mu$ route is related to the (weighted) average ${\overline{g}}(\eta)$ of the contact value $g(\eta;\xi)$ in the range $0\leq \xi\leq 1$, the virial route is directly related to the local value $g(\eta;1)$.  Since both  $g(\eta;\xi)$ and its first derivative $\partial g(\eta;\xi)/\partial \xi$ are  given exactly by the PY approximation at $\xi=0$ (see Fig.\ \ref{fig1}), it seems reasonable that the average
value ${\overline{g}}(\eta)$ is better estimated than the end point $g(\eta;1)$ by the PY approximation.
{In this respect, it is also worth noting that the weight function $(1+\xi)^{d-1}$ in Eq.\ \eqref{zmuf} increases with $\xi$ (except in the one-dimensional case, $d=1$), so that the average value ${\overline{g}}(\eta)$ is more influenced by the approximate values of $g(\eta;\xi)$ near $\xi=1$ than by the quasiexact values near $\xi=0$. This bias strongly increases with increasing dimensionality [in fact, $(1+\xi)^{d-1}\to (2^d/d)\delta(\xi-1)$ in the limit $d\to\infty$], which explains why the discrepancy between the PY $\mu$ and virial routes is much smaller with $d=5$ than with $d=3$.
}

%%%%%%%%%%%%%%%%%%%%%%%%%%%%%%%%%%%%%%%%
\section{Final remarks}\label{s.fr}

In this paper, we have considered an application of the Kirkwood coupling parameter method (or $\mu$ route) to the thermodynamics of classical fluids. Specifically, the EOS of the $5{\rm D}$-sphere fluid has been derived from the $\mu$ route by exploiting the knowledge of the exact solution of the PY theory for this system. The basic quantity required for this implementation is the contact value of the RDF for a partially coupled particle, which corresponds to a binary mixture with one infinitely dilute component.
Although the PY approximation had been solved for the $5{\rm D}$-sphere system by Freasier and Isbister \cite{FI81} and by Leutheusser \cite{L84}, their methods do not provide the RDF of a binary mixture required here. For this purpose we have used the RFA method \cite{RS11} which exactly solves the PY equation for hypersphere mixtures in odd dimensionalities. The RDF obtained in this way has an analytical expression and the resulting EOS in the $\mu$ route is derived from a numerical integration over density.

By examining the first few coefficients in the density expansion for the EOS and comparing them with available exact or  highly accurate values, it is seen that, although the PY theory is exact up to the third coefficient, it  rapidly worsens for increasing density order.
Specifically, the compressibility equation gives in  general the most satisfactory values, followed by the $\mu$ route, which becomes somewhat better than the virial one. Contrary to the case of $3{\rm D}$-sphere systems \cite{S12b}, linear interpolations of the PY EOS from different routes do not improve the evaluation of the virial coefficients.

Regarding the behavior at finite densities, the pressure predicted by the $\mu$ route is observed to lie below to simulation data within the fluid phase. Interestingly, the deviations of the compressibility and $\mu$ routes with respect to the numerical experiments are roughly of the same magnitude (but of opposite signs) in this region and both are lower than those from the virial route.
An accurate fit to simulation data in the fluid regime ($\eta<0.2$) is obtained if the  compressibility and $\mu$ routes are added together with equal weights. An alternative accurate fit to numerical experiments may be derived by a similar interpolation between the virial and compressibility routes with appropriate weights.
These results show a rather regular behavior of the three PY routes  around the exact EOS within the whole fluid region.

\acknowledgments
{The work of {R.D.R.} has been supported by the Consejo Nacional de Investigaciones Cient\'ificas y T\'ecnicas (CONICET, Argentina) through Grant No.\ PIP 114-201101-00208. A.S. acknowledges support from the Spanish government through Grant No.\ FIS2013-42840-P and by the Regional Government of Extremadura (Spain) through Grant No.\ GR15104 (partially financed by ERDF funds).}

\appendix
%%%%%%%%%%%%%%%%%%%%%%%%%%%%%%%%%%%%%

\section{RDF for a general five-dimensional binary mixture\label{appA}}

For an AHS binary mixture at $d=5$, the exact PY solution of the functional (\ref{Gdef}) may be expressed by \cite{RS11}
\beq \label{Grfa5}
G_{ij}(s) = \frac{e^{-\sigma_{ij}s}}{s^2}
\left[ (\sm{L}_0+\sm{L}_1 s+\sm{L}_2 s^2)\cdot\sm{B}^{-1}(s)\right]_{ij},
\eeq
where $\sm{L}_m$ ($m=0,1,2$) and $\sm{B}(s)$ are $2\times 2$  matrices. More specifically,
\beq \label{Bs}
\sm{B}(s)= \sm{I}+\rho\left[\sm{\Phi}_0(s)\cdot\sm{L}_0+\sm{\Phi}_1(s)
\cdot\sm{L}_1+\sm{\Phi}_2(s)\cdot\sm{L}_2\right],
\eeq
where $\sm{I}$ is the unit matrix and $\sm{\Phi}_m(s)$ ($m=0,1,2$) are diagonal matrices with elements
\beq \label{phimatrix}
[\sm{\Phi}_m(s)]_{ii} = {v_5} x_i \sigma_i^{{5-m}} \phi_{{5-m}}(\sigma_i s),
\eeq
where
\beq  \label{phi}
\phi_m(t) \equiv \frac 1{t^{m}}
\left[ \sum_{j=0}^m \frac{(-t)^j}{j!} -e^{-t} \right].
\eeq
Furthermore,
\beq \label{d5_L0}
\sm{L}_0= {3}\begin{bmatrix} 1 & 1 \\ 1 & 1 \\ \end{bmatrix},
\eeq
\beqn \label{d5_L1}
\sm{L}_1 &=& 3\begin{bmatrix} \sigma_{1} & \sigma_{12} \\ \sigma_{12} &
\sigma_{2} \\ \end{bmatrix} + \frac{3{\eta}}{1-6\eta}
 \left( \frac{2M_6}{M_5} \begin{bmatrix} 1 & 1 \\ 1 & 1 \\ \end{bmatrix} \right. \cr
 &&+ \left. 3
\begin{bmatrix} \sigma_1 & \sigma_2 \\ \sigma_1 & \sigma_2 \\ \end{bmatrix}
 - {\frac{10}{M_5}}  \begin{bmatrix} {x_1\sigma_1^4} & {x_2\sigma_2^4} \\ {x_1\sigma_1^4} & {x_2\sigma_2^4} \\ \end{bmatrix} \cdot\sm{L}_2 \right).
\eeqn
Here, $M_n\equiv x_1\sigma_1^n+x_2\sigma_2^n$ is the $n$th diameter moment and $\eta=v_5\rho M_5$ is the total packing fraction.
As for the  matrix $\sm{L}_2$, it obeys the quadratic equation
\beq \label{d5_L2}
 \sm{Q}_0 + \sm{Q}_1\cdot\sm{L}_2+\sm{Q}_{2}\cdot\sm{L}_{2}\cdot
\left( \sm{P}_{0}+\sm{P}_{1}\cdot\sm{L}_2 \right)=\mathsf{0},
\eeq
where
{
\beqa
\label{d5_QP}
\sm{P}_{0} &\equiv& \frac 12
\begin{bmatrix}
\sigma_1&0\\
0&\sigma_2\\
\end{bmatrix}
 +
\frac{2\eta}{M_5}
\begin{bmatrix}
x_1\sigma_1^6& x_1\sigma_1^6\\
x_2\sigma_2^6&x_2\sigma_2^6\\
\end{bmatrix}\nn
&&
 +\frac{3\eta/M_5}
{1-6\eta}
\begin{bmatrix}
x_1\sigma_1^6&x_1\sigma_1^5\sigma_2\\
x_2\sigma_2^5\sigma_1&x_2\sigma_2^6\\
\end{bmatrix}
\nn
&& + \frac{12 \eta^2 M_6/M_5^2}{1-6\eta}
\begin{bmatrix}
  x_1\sigma_1^5&x_1\sigma_1^5\\
  x_2\sigma_2^5&x_2\sigma_2^5\\
\end{bmatrix},
\eeqa
\beqa
\sm{P}_{1}& \equiv&-\frac {60\eta^2/M_5^2}{1-6\eta}
\begin{bmatrix}
  x_1^2\sigma_1^9&x_1x_2\sigma_1^5\sigma_2^4\\
  x_1x_2\sigma_1^4\sigma_2^5&x_2^2\sigma_2^9
\end{bmatrix}
\nn
&&
 -\frac{10\eta}{M_5}
\begin{bmatrix}
  x_1\sigma_1^4&0\\
  0&x_2\sigma_2^4\\
\end{bmatrix},
\eeqa
\beqa
\sm{Q}_0 &\equiv&- \frac{\eta}{M_5}\left[ \frac {20\eta^2M_6^3/M_5^2}{(1-6\eta)^2}+\frac{4\eta M_6M_7/M_5}{1-6\eta}
+\frac {M_8}{8} \right]
 \cr
&&\times \begin{bmatrix}
  1&1\\
  1&1\\
\end{bmatrix}
- \frac{\eta}{M_5}\left[\frac {10\eta M_6^2/ M_5}{(1-6\eta)^2}+\frac{M_7}{1-6\eta}\right]
\begin{bmatrix}
  \sigma_1&\sigma_{12}\\
  \sigma_{12}&\sigma_2\\
\end{bmatrix}
 \cr
&&- \frac {2\eta M_6/M_5}{ 1-6\eta}
\begin{bmatrix}
  \sigma_1^2&\sigma_{12}^2\\
  \sigma_{12}^2&\sigma_2^2\\
\end{bmatrix}
- \frac {3\eta/2}{1-6\eta}
\nn
&&
\times
\begin{bmatrix}
  \sigma_1^3&\sigma_1\sigma_2\sigma_{12}\\
 \sigma_1\sigma_2 \sigma_{12}&\sigma_2^3\\
\end{bmatrix}
- \eta\frac {(1+24\eta) M_6}{4 M_5(1-6\eta)^2}
\nn
&&
\times
\begin{bmatrix}
  \sigma_1^2&\sigma_{1}\sigma_2\\
  \sigma_{1}\sigma_2&\sigma_2^2\\
\end{bmatrix}
-\frac{1}{3}\begin{bmatrix}
  \sigma_1^3&\sigma_{12}^3\\
  \sigma_{12}^3&\sigma_2^3\\
\end{bmatrix},
\eeqa
\beqa
\sm{Q}_1 &\equiv& \frac {10\eta^2 M_6/ M_5^2}{1-6\eta}
\begin{bmatrix}
  x_1\sigma_1^5&x_2\sigma_2^5\\
  x_1\sigma_1^5&x_2\sigma_2^5\\
\end{bmatrix}
 +\frac{1}6
 \begin{bmatrix}
   \sigma_1&0\\
   0&\sigma_2\\
 \end{bmatrix}
\nn
&&+\frac {5\eta/2M_5}{1-6\eta}
\begin{bmatrix}
  x_1\sigma_1^6&x_2\sigma_2^5\sigma_1\\
  x_1\sigma_1^5\sigma_2&x_2\sigma_2^6\\
\end{bmatrix}
+\frac{2\eta}{3M_5}
\cr
&&
\times
\begin{bmatrix}
  x_1\sigma_1^6&x_2\sigma_2^6\\
  x_1\sigma_1^6&x_2\sigma_2^6\\
\end{bmatrix}
+ \frac{10\eta^2}{M_5^2}\left[\frac {10\eta M_6^2/ M_5}{(1-6\eta)^2}+\frac{ M_7}{1-6\eta}\right]
 \cr
&&
\times
\begin{bmatrix}
  x_1\sigma_1^4&x_2\sigma_2^4\\
  x_1\sigma_1^4&x_2\sigma_2^4\\
\end{bmatrix}
 + \frac {25\eta^2 M_6/ M_5^2}{(1-6\eta)^2}
\begin{bmatrix}
  x_1\sigma_1^5&x_2\sigma_2^4\sigma_1\\
  x_1\sigma_1^4\sigma_2&x_2\sigma_2^5\\
\end{bmatrix}\nn
&&
+ \frac {5\eta/2M_5}{1-6\eta}
\begin{bmatrix}
  x_1\sigma_1^6&x_2\sigma_2^4\sigma_1^2\\
  x_1\sigma_1^4\sigma_2^2&x_2\sigma_2^6\\
\end{bmatrix},
\eeqa
\beqa
\label{mQ2}
\sm{Q}_{2}& \equiv& \frac{1}{3}
\begin{bmatrix}
  1&0\\
  0&1\\
\end{bmatrix}
+ \frac{3\eta}{M_5}
\begin{bmatrix}
  x_1\sigma_1^5& x_2\sigma_2^5\\
   x_1\sigma_1^5& x_2\sigma_2^5\\
\end{bmatrix}
+ \frac {20\eta^2  M_6/M_5^2}{1-6\eta}\nn
&&\times
\begin{bmatrix}
  x_1\sigma_1^4& x_2\sigma_2^4\\
   x_1\sigma_1^4& x_2\sigma_2^4\\
\end{bmatrix}
+ \frac {5\eta/M_5}{1-6\eta}
\begin{bmatrix}
  x_1\sigma_1^5& x_2\sigma_2^4\sigma_1\\
   x_1\sigma_1^4\sigma_2& x_2\sigma_2^5\\
\end{bmatrix}.\nn
\eeqa
}

In general, the solution to Eq.\ (\ref{d5_L2}) must be obtained numerically \cite{RS11}. However, an analytical form is possible for a binary mixture with an infinitely diluted component (the solute), as shown in Appendix \ref{appB}.

The asymptotic behavior of the functional $G_{ij}(s)$ required in the evaluation of the RDF contact value [Eq.\ (\ref{gG})] can be derived from
(\ref{Grfa5}) as
\beq \label{Gsgde}
\sigma_{ij}^2g_{ij}(\sigma_{ij}^+)=\lim_{s\rightarrow\infty}e^{\sigma_{ij}s}G_{ij}(s)=
 \left[\sm{L}_{2}\cdot\sm{B}^{-1}_\infty\right]_{ij},
\eeq
{where we have called $\sm{B}_\infty\equiv \lim_{s\to\infty}\sm{B}(s)$.}
 It was shown in Ref.\ \cite{RS11} that
\beq \label{Bsd5}
\sm{B}_\infty= \sm{I} {-}\rho \left(\sm{C}_5 \cdot\sm{L}_0
  +\sm{C}_4 \cdot\sm{L}_1 +\sm{C}_3 \cdot\sm{L}_2\right),
\eeq
where{
\beq
\sm{C}_m=(-1)^{m+1}\frac{(2\pi)^2}{m!}\begin{bmatrix} x_1\sigma_1^m & 0 \\ 0 & x_2\sigma_2^m \\ \end{bmatrix}.
\label{Cm}
\eeq
}

\section{Test particle limit \label{appB}}
The expressions in Appendix \ref{appA} apply for any five-dimensional AHS binary mixture in the PY approximation. Now, without loss of  generality, we adopt the convention $i=1$ and $i=2$ for solvent and solute particles, respectively, and denote by $\sigma_1=\sigma$ and $\sigma_2=\xi\sigma_1$ the respective diameters. Next, we take the test particle limit for the solute, i.e., $x_1\to 1$, $x_2\to 0$, so that $M_n\to \sigma^n$. This drastically simplifies the matrices \eqref{d5_QP}--\eqref{mQ2}.

Let us write the matrix $\sm{L}_2$ as
\beq \label{L2P0}
\sm{L}_2=\sigma^2
\begin{bmatrix} \lambda_{11} &\lambda_{12} \\\lambda_{21} & \lambda_{22} \\ \end{bmatrix}.
\eeq
The $(1,1)$ element of Eq.\ \eqref{d5_L2} yields a closed quadratic equation for $\lambda_{11}$. The physical solution  is identified by the condition $\lim_{\eta\to 0}\lambda_{11}=\text{finite}$ with the result
\beq
\label{a.e1}
\lambda_{11}={\frac{1+4\eta}{20\eta}-\frac{(1-\eta)(1-6\eta)}{20\eta\zeta}},
\eeq
where $\zeta$ is defined by Eq.\ \eqref{a.xi}.
Since $\lambda_{11}$ is a coefficient related to the bulk fluid, it does not depend on $\xi$ and its  expression  is equivalent to that of the one-component fluid [as given by Eq.\ (E10) of Ref.\ \cite{RS07}]. As for the remaining elements of the matrix $\sm{L}_2$, they are given from Eq.\ \eqref{d5_L2} by
{
\beq
\lambda_{12}=\frac{a_{12}/ 4 \zeta^2- 2 \eta  (2 + 3 \xi)\lambda_{11}}{
  1 + \xi + 2 \eta (2 - 10 \lambda_{11} - 3 \xi)},
    \eeq
\beq
\lambda_{21}=\frac{a_{12}-4  \eta b_{21} \lambda_{11} + 240 (3 + 5 \xi + 2 \eta) \eta^2 \lambda_{11}^2}{4 (1 - 6 \eta)
 \left[ 1 + \xi + 2 \eta (2 - 10 \lambda_{11} - 3 \xi)\right]},
    \eeq
\beq
\lambda_{22}=\frac{a_{22} -
   8 \eta   b_{22}\lambda_{12}-
   8 \eta (2 + 3 \xi - 10 \lambda_{12}) c_{22} }{8 (1 - 6 \eta)^2 \xi},
   \eeq
where
\bal
a_{12}=&(1 + \xi)^3 +
    3 \eta (5 + 8 \xi - 2 \xi^2 - 4 \xi^3)  \nn
    &+
    12 \eta^2 (6 + 4 \xi - 6 \xi^2 + 3 \xi^3) + 12 \eta^3,
    \eal
\beq
b_{21}=13 + 30 \xi + 15 \xi^2 + 6 \eta (19 + 20 \xi - 15 \xi^2) + 48 \eta^2,
\eeq
\bal
a_{22}=&8 \xi^3 + 3 \eta (1 + 8 \xi + 18 \xi^2 - 20 \xi^3) \nn
&+
 12 \eta^2 (5 + 8 \xi - 12 \xi^2 + 6 \xi^3) +
 12 \eta^3,
\eal
\beq
b_{22}=2 + 12 \xi + 15 \xi^2 + 6 \eta (6 + \xi - 15 \xi^2) + 6 \eta^2 (2 - 3 \xi),
\eeq
\beq
   c_{22}= 3 \eta  (3 + 5 \xi + 2 \eta)\lambda_{11}+(1 - 6 \eta) \lambda_{21}.
\eeq
}
It can be easily  checked that if the solute is equivalent to a solvent particle (i.e., $\xi=1$), then the elements $\lambda_{12}$, $\lambda_{21}$, and $\lambda_{22}$ coincide with $\lambda_{11}$, as expected.

Now we determine the contact values. Taking into account that $\eta=\pi^2\rho\sigma^5/60$ and using Eqs.\ \eqref{d5_L0}, \eqref{d5_L1},  \eqref{Cm}, and \eqref{L2P0}, one obtains
\beq
\rho \sm{C}_5\cdot\sm{L}_0 = 6\eta
\begin{bmatrix} 1 & 1 \\ 0 & 0 \\ \end{bmatrix},\quad
\rho \sm{C}_3\cdot\sm{L}_2 = 40\eta
\begin{bmatrix} \lambda_{11} &\lambda_{12} \\ 0 & 0 \\ \end{bmatrix},
\eeq
\beq
\rho \sm{C}_4\cdot\sm{L}_1 = {-{30\eta}
\begin{bmatrix} \frac{1-\eta(1+10\lambda_{11})}{1-6\eta} & \frac{{1+\xi}-2\eta(1+10\lambda_{12})}{2(1-6\eta)} \\ 0 & 0 \\ \end{bmatrix}}.
\eeq
Insertion of this into Eq.\ \eqref{Bsd5} yields
\beq
\sm{B}_\infty=
\begin{bmatrix}
\frac{\zeta^2-20\eta(2+3\eta)\lambda_{11}}{1-6\eta} &
\frac{3\eta(3+2\eta+5\xi)-20\eta(2+3\eta)\lambda_{12}}{1-6\eta} \\
0 & 1 \\
\end{bmatrix}.
\eeq
Its inverse matrix is
\beq
\sm{B}^{-1}_\infty=
\begin{bmatrix}
\frac{1-6\eta}{\zeta^2-20\eta(2+3\eta)\lambda_{11}} &
\frac{20\eta(2+3\eta)\lambda_{12}-3\eta(3+2\eta+5\xi)}
{\zeta^2-20\eta(2+3\eta)\lambda_{11}} \\
0 & 1 \\
\end{bmatrix}.
\eeq
Finally, from (\ref{Gsgde}) one has
{
\beq
g_{11}(\sigma_1^+)=\frac{1-6\eta}{\zeta^2/\lambda_{11}-20\eta(2+3\eta)},
\label{g11cA}
\eeq
\bal
g_{12}(\sigma_{12}^+)=&\frac{4}{\left(1+\xi\right)^2}\Big[\frac{20\eta(2+3\eta)\lambda_{12}-3\eta(3+2\eta+5\xi)}
{\zeta^2/\lambda_{11}-20\eta(2+3\eta)}\nn
&+\lambda_{12}\Big],
\label{g12cA}
\eal
\beq
g_{21}(\sigma_{12}^+)=\frac{4}{\left(1+\xi\right)^2}\frac{1-6\eta}{\zeta^2/\lambda_{11}-20\eta(2+3\eta)}\frac{\lambda_{21}}{\lambda_{11}},
\eeq
\bal
g_{22}(\sigma_{2}^+)=&\frac{1}{\xi^2}\Big[\frac{20\eta(2+3\eta)\lambda_{12}-3\eta(3+2\eta+5\xi)}
{\zeta^2/\lambda_{11}-20\eta(2+3\eta)}\frac{\lambda_{21}}{\lambda_{11}}\nn
&+\lambda_{22}\Big].
\eal
}
As expected, $g_{12}(\sigma_{12}^+)=g_{21}(\sigma_{12}^+)$. By inserting the explicit expressions of $\lambda_{11}$ and $\lambda_{12}$ it is straightforward to check that Eqs.\ \eqref{g11cA} and \eqref{g12cA} reduce to Eq.\ \eqref{g11c} and \eqref{g12c}, respectively.

\bibliographystyle{apsrev}\bibliography{D:/Dropbox/Public/bib_files/liquid}
\end{document}